\documentclass[aps,pra,twoside,superscriptaddress,twocolumn,showpacs,floatfix]{revtex4}

\usepackage{latexsym}
\usepackage{graphicx}

\usepackage{amssymb}
\usepackage{amsmath}

\usepackage{color}

\begin{document}

\title{Optimal entanglement-assisted discrimination of quantum measurements}

\author{M. Mikov\'{a}}
\affiliation{Department of Optics, Palack\'{y} University, 17. listopadu 1192/12, CZ-771 46 Olomouc, Czech Republic}

\author{M. Sedl\'{a}k}
\affiliation{Department of Optics, Palack\'{y} University, 17. listopadu 1192/12, CZ-771 46 Olomouc, Czech Republic}
\affiliation{RCQI,~Institute of Physics,~Slovak Academy of Sciences,~D\'ubravsk\'a cesta 9,~84511 Bratislava,~Slovakia}

\author{I. Straka}
\affiliation{Department of Optics, Palack\'{y} University, 17. listopadu 1192/12, CZ-771 46 Olomouc, Czech Republic}

\author{M. Mi\v{c}uda}
\affiliation{Department of Optics, Palack\'{y} University, 17. listopadu 1192/12, CZ-771 46 Olomouc, Czech Republic}

\author{M. Ziman}
\affiliation{RCQI,~Institute of Physics,~Slovak Academy of Sciences,~D\'ubravsk\'a cesta 9,~84511 Bratislava,~Slovakia}
\affiliation{Faculty of Informatics,~Masaryk University,~Botanick\'a 68a,~60200 Brno,~Czech Republic}

\author{M. Je\v{z}ek}
\affiliation{Department of Optics, Palack\'{y} University, 17. listopadu 1192/12, CZ-771 46 Olomouc, Czech Republic}

\author{M. Du\v{s}ek}
\affiliation{Department of Optics, Palack\'{y} University, 17. listopadu 1192/12, CZ-771 46 Olomouc, Czech Republic}

\author{J. Fiur\'{a}\v{s}ek}
\affiliation{Department of Optics, Palack\'{y} University, 17. listopadu 1192/12, CZ-771 46 Olomouc, Czech Republic}

\begin{abstract}
We investigate optimal discrimination between two projective single-qubit measurements in a scenario where the measurement can be performed only once. We consider general setting
involving a tunable fraction of inconclusive outcomes and we prove that the optimal discrimination strategy requires an entangled probe state for any nonzero rate of inconclusive outcomes.
 We experimentally implement this optimal discrimination strategy for projective measurements on polarization states of single photons. Our setup
 involves a real-time electrooptical feed-forward loop which allows us to fully harness the benefits of entanglement in discrimination of quantum measurements. The experimental data clearly demonstrate
 the advantage of entanglement-based discrimination strategy as compared to  unentangled single-qubit probes.
\end{abstract}


\maketitle

\section{Introduction}

One of the characteristic traits of quantum mechanics is the impossibility to perfectly discriminate two non-orthogonal quantum states. This fundamental property
of quantum systems has far reaching practical implications ranging from security of quantum key distribution protocols to limits on measurement precision in metrologic schemes.
Impossibility of perfect discrimination also immediately triggers the question what is the optimal approximate or probabilistic discrimination strategy.
Given their wide range of potential applications, such strategies have been studied in great detail both theoretically \cite{helstrom,idp1,idp4,Chefles98,Zhang99,intmdt0,intmdteq1,intmdt1,intmdt2,usd_qkd,prog_discr}
and experimentally \cite{expidp1,exphelstrom,expidp2,discr_coh,prog_discr_ex}.
More recently, this concept has been extended to discrimination of quantum operations \cite{acin,dariano,sacchi,wang,duan,piani,watrous,ziman1,hashimoto,obrien,expzhang,dallarno}
and measurements \cite{ziman2,mdiscr1,mdiscr2}.
While sharing many similarities with discrimination of quantum states, discrimination of quantum devices admits intriguing
novel strategies and phenomena \cite{zimanppovm,memeff,architecture,supermaps,comblong,gutoski,chiribella} such as using probes entangled with auxiliary systems, or the perfect distinguishability
of any two unitary operations when a sufficiently large but finite number of copies of the operation is available \cite{acin}.

\begin{figure}[!b!]
\centerline{\includegraphics[width=0.98\linewidth]{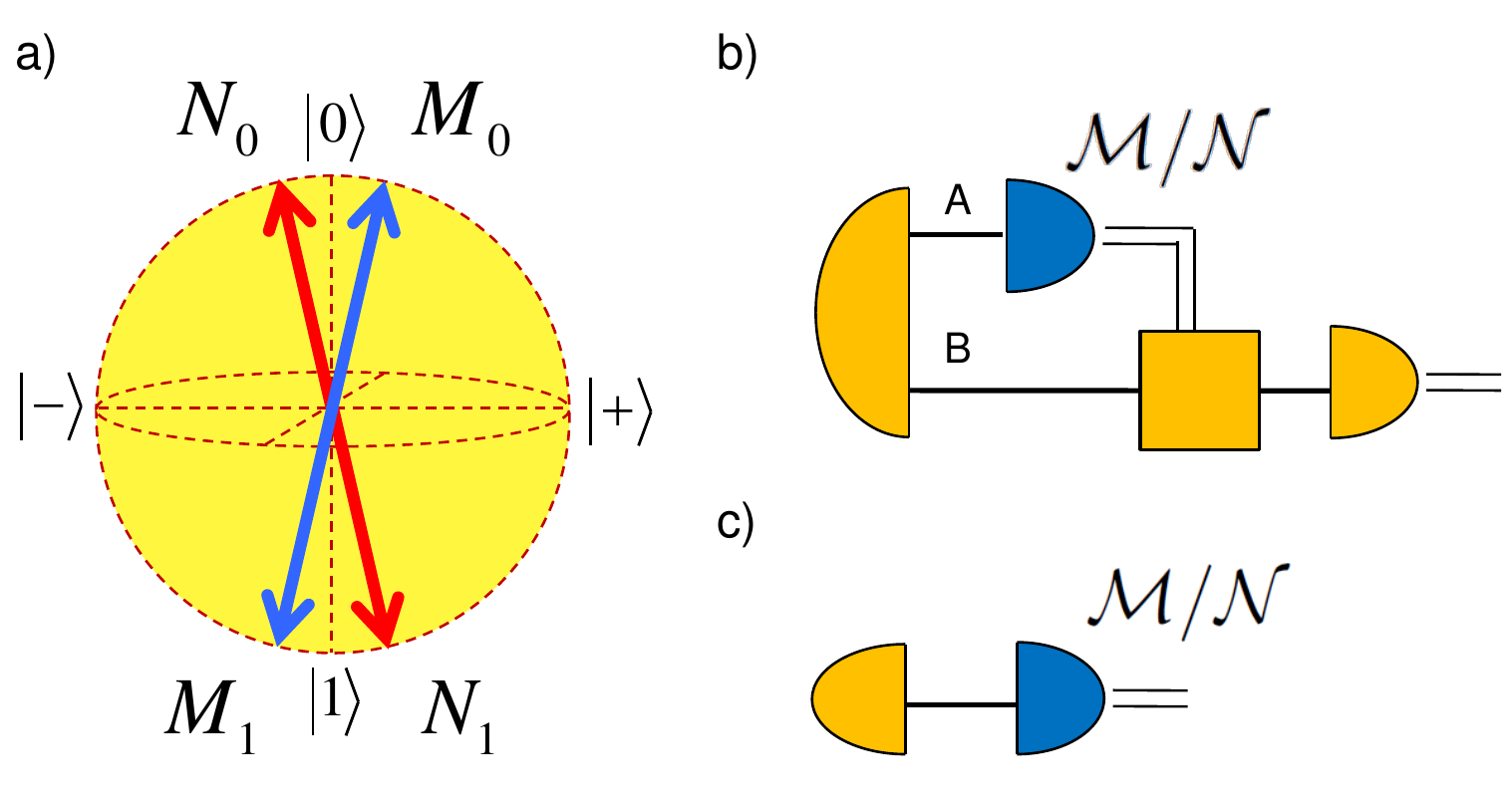}}
\caption{(Color online) (a) Single-qubit measurements $\mathcal{M}$ and $\mathcal{N}$ on a Bloch sphere. (b) General measurement discrimination scheme involving entangled probe state.
(c) Simple discrimination scheme with single-qubit probe.}
\end{figure}

Here, we investigate the utility of entanglement for the canonical task of optimal discrimination between two projective measurements $\mathcal{M}$ and $\mathcal{N}$ on a single qubit
provided that the measurement can be performed only once.
We consider general discrimination strategies involving a certain fraction of inconclusive
outcomes, $P_I$, and we show that the optimal discrimination procedure requires entangled probe state
unless $P_I=0$. As a benchmark, we also provide the optimal discrimination scheme with no entanglement. We experimentally implement the optimal discrimination
for projective measurements on polarization states of single photons. Our setup is based on linear optics, real-time feed-forward-loop, fiber interferometers,
and single-photon detectors. Experimental data unequivocally confirm the advantage of entanglement-based
discrimination strategies.

\section{Optimal entanglement-assisted discrimination}

The measurement bases $\mathcal{M}$ and $\mathcal{N}$ are illustrated in Fig.~1(a). Without loss of generality, the projectors specifying the measurements can be parameterized by a single angle $\theta$,
\begin{equation}
\begin{array}{lll}
M_0=|\phi\rangle\langle \phi|, &\quad & M_1=|\phi^\perp\rangle\langle \phi^\perp|, \\[1mm]
N_0=|\psi\rangle \langle \psi|, & \quad & N_1=|\psi^\perp\rangle \langle \psi^\perp|,
\end{array}
\label{MNdefinition}
\end{equation}
where
\begin{equation}
\begin{array}{lll}
|\phi\rangle=\cos\theta|0\rangle+\sin\theta|1\rangle, & & |\phi^\perp\rangle=\sin\theta|0\rangle-\cos\theta|1\rangle, \\[1mm]
|\psi\rangle=\cos\theta|0\rangle-\sin\theta|1\rangle, & & |\psi^\perp\rangle=\sin\theta|0\rangle+\cos\theta|1\rangle,
\end{array}
\label{phipsidefinition}
\end{equation}
and $0\leq \theta \leq \frac{\pi}{4}$.
The most general discrimination strategy is depicted in Fig. 1(b). A two-qubit entangled state  $|\Psi\rangle_{AB}$  is employed, the measurement that should be identified is performed on qubit A,
and the measurement outcome ($0$ or $1$) specifies which measurement is then performed on the other qubit $B$.

In what follows we assume equal a-priori probabilities of the two measurements. In such a case we will show it is optimal to employ a maximally
entangled singlet Bell state $|\Psi^{-}\rangle=(|01\rangle-|10\rangle)/\sqrt{2}$.
If we observe measurement outcome $0$ on qubit $A$, then qubit $B$ is prepared in the state $|\phi^\perp\rangle$ or $|\psi^\perp\rangle$.
Similarly, outcome $1$ heralds that qubit B is prepared in the state $|\phi\rangle$ or $|\psi\rangle$. The discrimination of quantum measurements is in this way converted to discrimination of two non-orthogonal quantum states.
Since
\begin{equation}
|\phi\rangle=-\sigma_Y|\phi^\perp\rangle, \quad |\psi\rangle=\sigma_Y|\psi^\perp\rangle,
\label{sigmaYconnection}
\end{equation}
we can apply the unitary operation $\sigma_Y=|0\rangle \langle 1|-|1\rangle \langle 0|$
to qubit $B$ when the measurement outcome on $A$
reads $0$, and we end up with the task to discriminate between two fixed non-orthogonal states $|\phi\rangle$ and $|\psi\rangle$.

As shown by Ivanovic, Dieks, and Peres (IDP) \cite{idp1}, perfect error-free discrimination between $|\phi\rangle$ and $|\psi\rangle$ is possible if we
allow for a certain probability of inconclusive outcomes $P_I=|\langle \psi|\phi\rangle|$. Explicitly, we have
 $P_I=\cos(2\theta)$.
Unambiguous discrimination requires a generalized $3$-component POVM which can be interpreted as a quantum filtering followed by projective measurement on the filtered state.
The required filter has the form $F=\tan\theta|0\rangle \langle 0| + |1\rangle\langle 1|$ and the filtered states become orthogonal, $F|\phi\rangle=\sqrt{2}\sin\theta |+\rangle$, and
$F|\psi\rangle=\sqrt{2}\sin\theta |-\rangle$, where $|\pm\rangle=(|0\rangle\pm|1\rangle)/\sqrt{2}$. The square of the
 norm of the filtered states is equal to the success probability of unambiguous discrimination, $P_S=2\sin^2\theta$, and $P_S+P_I=1$.

Due to the various experimental imperfections, we will in practice encounter also erroneous conclusive results occurring with probability $P_E$. This motivates us to consider a general discrimination scheme
where we maximize $P_S$ (hence minimize $P_E$) for a fixed fraction of inconclusive outcomes $P_I$. The optimal filter then reads $F=f|0\rangle\langle 0|+|1\rangle\langle 1|$, where $f=\sqrt{1-P_I/\cos^2\theta}$,
and a projective measurement in basis $|\pm\rangle$ should be performed after successful filtration similarly as before. This intermediate strategy optimally interpolates
between IDP \cite{idp1} and Helstrom \cite{helstrom} schemes, and we get \cite{Chefles98,Zhang99}
\begin{equation}
P_S=\frac{1}{2}\left(1-P_I+\sin(2\theta)\sqrt{1-\frac{P_I}{\cos^2\theta}}\right).
\label{PSentangled}
\end{equation}
It is convenient to consider also a relative probability of  successful discrimination for the subset of conclusive outcomes,
 $\tilde{P}_S=P_S/(1-P_I).$
$\tilde{P}_S$ increases with $P_I$ and $\tilde{P}_S=1$ when $P_I=\cos(2\theta)$.

The optimality of the above protocol can be proved with the help of the formalism of process POVM \cite{zimanppovm,memeff}.
We associate $i$th output of the measurement device with quantum state $|i\rangle$, $i=1,0$, and associate measurement $X$
with operator $E_X=X^T_0\otimes |0\rangle\langle 0|+X^T_1\otimes |1\rangle\langle 1|$, where $X\in \{M,N\}$. An arbitrary test that discriminates
between the measurements $\mathcal{M}$ and $\mathcal{N}$ and is allowed by quantum mechanics
is described by a 3-component process POVM  $\{T_M,T_N,T_I\}$ on a Hilbert space of two qubits, where $T_k \geq 0$ and $T_M+T_N+T_I=\rho \otimes \mathbb{I}$. Here $\rho$ denotes a density matrix
of a single qubit, $\rho\geq 0$ and $\mathrm{Tr}[\rho]=1$, and $\mathbb{I}$ represents an identity operator.
Results $T_M$ and $T_{N}$ correspond to guessing measurement $\mathcal{M}$ and $\mathcal{N}$, respectively, while $T_I$
represents the inconclusive outcomes. Within this formalism, the probabilities $P_S$, $P_E$ and $P_I$ can be expressed as follows,
\begin{eqnarray}
P_S&=& \frac{1}{2}\left(\mathrm{Tr}[T_M E^T_M]+\mathrm{Tr}[T_N E^T_N]\right), \nonumber \\
P_E&=& \frac{1}{2}\left(\mathrm{Tr}[T_M E^T_N]+\mathrm{Tr}[T_N E^T_M]\right), \nonumber \\
P_I&=&\frac{1}{2}\mathrm{Tr}\left[T_I(E^T_M+E^T_N)\right].
\label{Pexpressions}
\end{eqnarray}
Thanks to the block-diagonal structure of $E_M$ and $E_N$ it suffices to consider $T_k=H_{k,0}\otimes|0\rangle\langle 0|+ H_{k,1}\otimes|1\rangle\langle 1|$ and the constraint on $T_k$ can be rephrased as
\begin{equation}
H_{M,i}+H_{N,i}+H_{I,i}=\rho, \qquad i=0,1.
\label{Hconstraint}
\end{equation}
Furthermore, due to the property (\ref{sigmaYconnection}) it suffices to consider only covariant $T_k$, where $H_{k,1}=\sigma_Y H_{k,0}\sigma_Y^\dagger$ and $\rho=\sigma_Y \rho \sigma_Y^\dagger$.
This can be seen by noting that the following substitutions do not alter the value of probabilities (\ref{Pexpressions}) while making  $T_k$ covariant,
\begin{equation}
\begin{array}{l}
H_{k,0}\rightarrow \frac{1}{2}(H_{k,0}+\sigma_Y H_{k,1}\sigma_Y^\dagger), \\[1mm]
 H_{k,1}\rightarrow \frac{1}{2}(H_{k,1}+\sigma_Y H_{k,0}\sigma_Y^\dagger).
\end{array}
\label{Hcovariant}
\end{equation}
Finally, since the projectors (\ref{MNdefinition}) are real, one can also choose $H_{k,i}$ to be real and set their imaginary parts to zero without
changing the probabilities (\ref{Pexpressions}). This means that $\rho$ is real as well, which
together with $\rho=\sigma_Y\rho\sigma_Y^\dagger$ implies that $\rho=\mathbb{I}/2$. If we combine together all the above results, we find that the probabilities (\ref{Pexpressions}) can be expressed as
\begin{eqnarray}
P_{S}&=&\mathrm{Tr}[H_{M,0} M_0]+ \mathrm{Tr}[H_{N,0} N_0], \nonumber \\
P_{E}&=&\mathrm{Tr}[H_{N,0} M_0] + \mathrm{Tr}[H_{M,0} N_0] , \nonumber \\
P_{I}&=&\mathrm{Tr}[H_{I,0}(M_0+N_0)],
\label{PHexpressions}
\end{eqnarray}
and the operators $H_{k,0}$ satisfy the conditions $H_{k,0}\geq 0$, and $H_{M,0}+H_{N,0}+H_{I,0}=\mathbb{I}/2$. This shows that the optimization of
discrimination of two projective qubit measurements becomes equivalent to optimization of the
discrimination of two quantum states $M_0$ and $N_0$ by a $3$-component POVM with elements $2H_{M,0}$, $2H_{N,0}$, and $2H_{I,0}$.

\section{Optimal discrimination with single-qubit probes}

To elucidate the importance of entanglement for measurement discrimination and to provide a benchmark for the experiment, we now determine the optimal discrimination strategy with unentangled single-qubit
probes, see Fig. 1(c). In this case one has to guess $\mathcal{M}$ or $\mathcal{N}$ solely based on the measurement outcome on the probe qubit. We shall show that the optimal strategy for a fixed probe state
can be constructed such that for observation $0$ we always guess $\mathcal{M}$  while for observation $1$ we guess $\mathcal{N}$ with probability $q$ and provide an inconclusive outcome with probability $1-q$. Let $\rho$ denote
density matrix of the probe state and define $P_{M,i}=\mathrm{Tr}[M_i \rho]$, $P_{N,i}=\mathrm{Tr}[N_i \rho]$. We can always re-label the measurements and outcomes such that
\begin{equation}
\frac{P_{M,0}}{P_{N,0}}\geq\frac{P_{N,1}}{P_{M,1}}\geq 1.
\label{Pinequality}
\end{equation}
Note that $P_{M,0} \geq P_{N,0}$ implies $P_{N,1}\geq P_{M,1}$ because $P_{M,0}+P_{M,1}=P_{N,0}+P_{N,1}=1$.
First observe that it does not help to produce inconclusive outcomes for both observations $0$ and $1$, because this only increases $P_I$ while not further improving $\tilde{P}_S$ with respect to the strategy where
inconclusive results are declared only for outcome $1$. The inequalities (\ref{Pinequality}) then imply the optimality of the above defined strategy and we can write
\begin{eqnarray}
P_{S}&=&\frac{1}{2}\left(\mathrm{Tr}[M_{0}\rho]+q\mathrm{Tr}[N_{1}\rho]\right),  \nonumber \\
P_{I}&=&\frac{1}{2}(1-q)\mathrm{Tr}[(M_1+N_1)\rho],
\label{PSIsingle}
\end{eqnarray}
and $P_{E}=1-P_S-P_I$.
It is easy to verify that for a fixed $P_I$ the probability $P_S$ is maximized when the probe state is pure with
 real amplitudes, $\rho=|\vartheta\rangle\langle \vartheta|$, where $|\vartheta\rangle=\cos\vartheta|0\rangle+\sin\vartheta|1\rangle$.
Explicitly, we get
\begin{eqnarray}
P_S&=&\frac{1}{2}\left[1+\sin(2\theta)\sin(2\vartheta)-(1-q)\sin^2(\theta+\vartheta)\right],
\nonumber \\
P_I&=&\frac{1-q}{2}(1-cx),
\label{PSIdef}
\end{eqnarray}
where  $c=\cos(2\theta)$ and $x=\cos(2\vartheta)$.

Using Eq. (\ref{PSIdef}) we can express $q$ as a function of $P_I$,
\begin{equation}
q=1-\frac{2P_I}{1-xc}.
\end{equation}
If we insert this formula for $q$ into Eq. (\ref{PSIdef}), we obtain
\begin{equation}
P_S=\frac{1}{2}(1-P_I)+\frac{1}{2}\sqrt{(1-c^2)(1-x^2)}\left[1-\frac{P_I}{1-xc}\right].
\label{PS}
\end{equation}
The optimal $\vartheta$ that maximizes $P_S$ for a given $P_I$ can be determined from the condition
\begin{equation}
\frac{\partial P_S}{\partial x}=0,
\end{equation}
which leads to a qubic equation for $x$,
\begin{equation}
c^2 x^3-2cx^2+(1-P_I)x+P_Ic=0.
\label{xcubic}
\end{equation}
This construction is applicable only if $q>0$, which is equivalent to  $P_{I}<P_{I,B}$, where the boundary $P_{I,B}$ can be determined from the condition that $x$ satisfies
Eq. (\ref{xcubic}) and, simultaneously, $q=0$. After some algebra, this yields a quadratic equation $8P_{I,B}^2-6P_{I,B}+1-c^2=0,$
 whose solution reads
\begin{equation}
P_{I,B}=\frac{1}{8}\left(3+\sqrt{1+8 c^2}\right),
\label{PIB}
\end{equation}
If $P_I \geq P_{I,B}$, then it is optimal to set $q=0$. This implies $x=(1-2P_I)/c$
and
\begin{equation}
P_S=\frac{1}{2}(1-P_I)+\frac{1}{4}\sin(2\theta)\sqrt{1-\frac{(1-2P_I)^2}{\cos^2(2\theta)}}.
\label{PSsingle}
\end{equation}
Explicit numerical calculations reveal that the resulting dependence of $P_S$ on $P_I$ is a convex function for $P_I<P_{I,B}$, see Appendix.
Eq. (\ref{PS}) therefore does not determine the optimal discrimination
strategy with single-qubit probes.
The situation is depicted in Fig.~2.
The crosses represent the dependence of $P_S$ on $P_I$ specified by
Eqs.~(\ref{PS}) and (\ref{PSsingle}). Since $P_S$ is a convex function of $P_I$ for $P_I<P_{I,B}$,
the area below the curve $P_S(P_I)$ does not form a convex set.

\begin{figure}[!t!]
\includegraphics[width=0.95\linewidth]{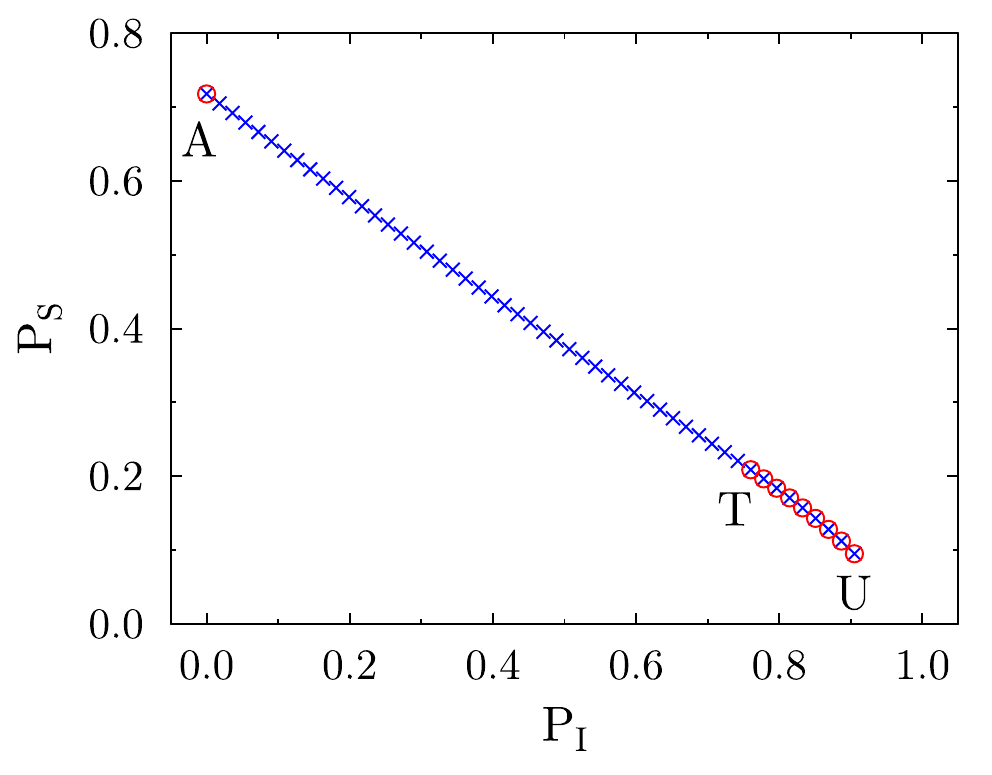}
\caption{Blue crosses show the dependence of $P_S$ on $P_I$ as specified by Eqs. (\ref{PS}) and (\ref{PSsingle}), $c=0.9$. The red circles indicate the convex hull,
points $A$ and $U$ correspond to minimum error and unambiguous discrimination with single-qubit probes, respectively, and point $T$ is specified by Eq. (\ref{PITdefinition}).}
\end{figure}

In order to obtain the optimal discrimination strategy with single-qubit probes, we must construct a convex hull of the discrimination strategies represented by blue crosses in Fig.~2.
The result is indicated by red circles. Geometrically, we must construct a tangent line to the curve specified by Eq. (\ref{PSsingle}), which passes through the point $A$
that corresponds to the optimal minimum error discrimination:  $\vartheta=\pi/4$, $P_{I,0}=0$, $P_{S,0}=[1+\sin(2\theta)]/2$.
This tangent line touches the curve (\ref{PSsingle}) at point $T$, which is specified by
\begin{equation}
P_{I,T}=\frac{1+3c^2+2c^2\sqrt{1+3c^2}}{2(1+4c^2)}.
\label{PITdefinition}
\end{equation}
Note that $P_{I,T} \geq P_{I,B}$.
In the interval $0<P_I< P_{I,T}$  the optimal discrimination strategy is thus a mixture of two strategies corresponding to points $A$ and $T$
with weights $1-P_I/P_{I,T}$ and $P_I/P_{I,T}$, respectively. This means, that with probability $1-P_I/P_{I,T}$ we should perform
the optimal minimum-error discrimination with probe state $|\vartheta\rangle=|+\rangle$ and $q=1$, which results in $P_{S,0}=[1+\sin(2\theta)]/2$ and $P_{I,0}=0$.
With probability $P_I/{P_{I,T}}$ we should use the probe state with $x=(1-2P_{I,T})/c$, which yields $P_S=P_{S,T}$ given by Eq. (\ref{PSsingle}),
where $P_I$ is replaced with $P_{I,T}$. The overall success probability then reads
\begin{equation}
P_{S}=\left(1-\frac{P_{I}}{P_{I,T}}\right) P_{S,0}+ \frac{P_I}{P_{I,T}}P_{S,T}.
\label{PSsinglefinal}
\end{equation}
If $P_{I}\geq P_{I,T}$, then it is optimal to use only one single-qubit probe specified by $x=(1-2P_I)/c$.
In this case, the optimal $P_S$ is given by Eq. (\ref{PSsingle}), see also the circles in Fig.~2. The end-point $U$ corresponds to unambiguous
discrimination with a single-qubit probe: $\vartheta=\pi/2-\theta$, $P_S=(1-c^2)/2$, and $P_I=(1+c^2)/2$.

To verify the validity of our analytical construction, we have performed extensive numerical analysis of the convex hulls for various
values of $c$ using the MATLAB function $\mathrm{convhull}$. For each chosen $c$, we have generated $10^4$ pairs $(P_I,P_S)$ corresponding
to discrimination strategies described by Eqs. (\ref{PS}) and (\ref{PSsingle}), and we have numerically calculated the convex hull.
In all cases, the convex hull constructed in this way had the structure illustrated in Fig.~2 and the position of point $T$ agreed
with the analytical formula (\ref{PITdefinition}).


\section{Experiment}

Our experimental demonstration of entanglement-assisted discrimination of quantum measurements is based on linear optics and qubits encoded into states of single photons.
The scheme of our experimental setup is shown in Fig.~3.
Time-correlated orthogonally polarized photon pairs were generated
by the process of collinear frequency-degenerate type-II spontaneous parametric down-conversion in a 2 mm thick BBO crystal pumped by a CW laser diode at 405~nm.
A post-selected two-photon polarization singlet Bell state $|\Psi^{-}\rangle$ was prepared by interfering the vertically polarized signal photon
and horizontally polarized idler photon at a balanced beam splitter (BS). The state was characterized by quantum state tomography
and we observed purity $>98\%$ and fidelity $>99\%$.

\begin{figure} [!t!]
\centerline{\includegraphics[width=0.98\linewidth]{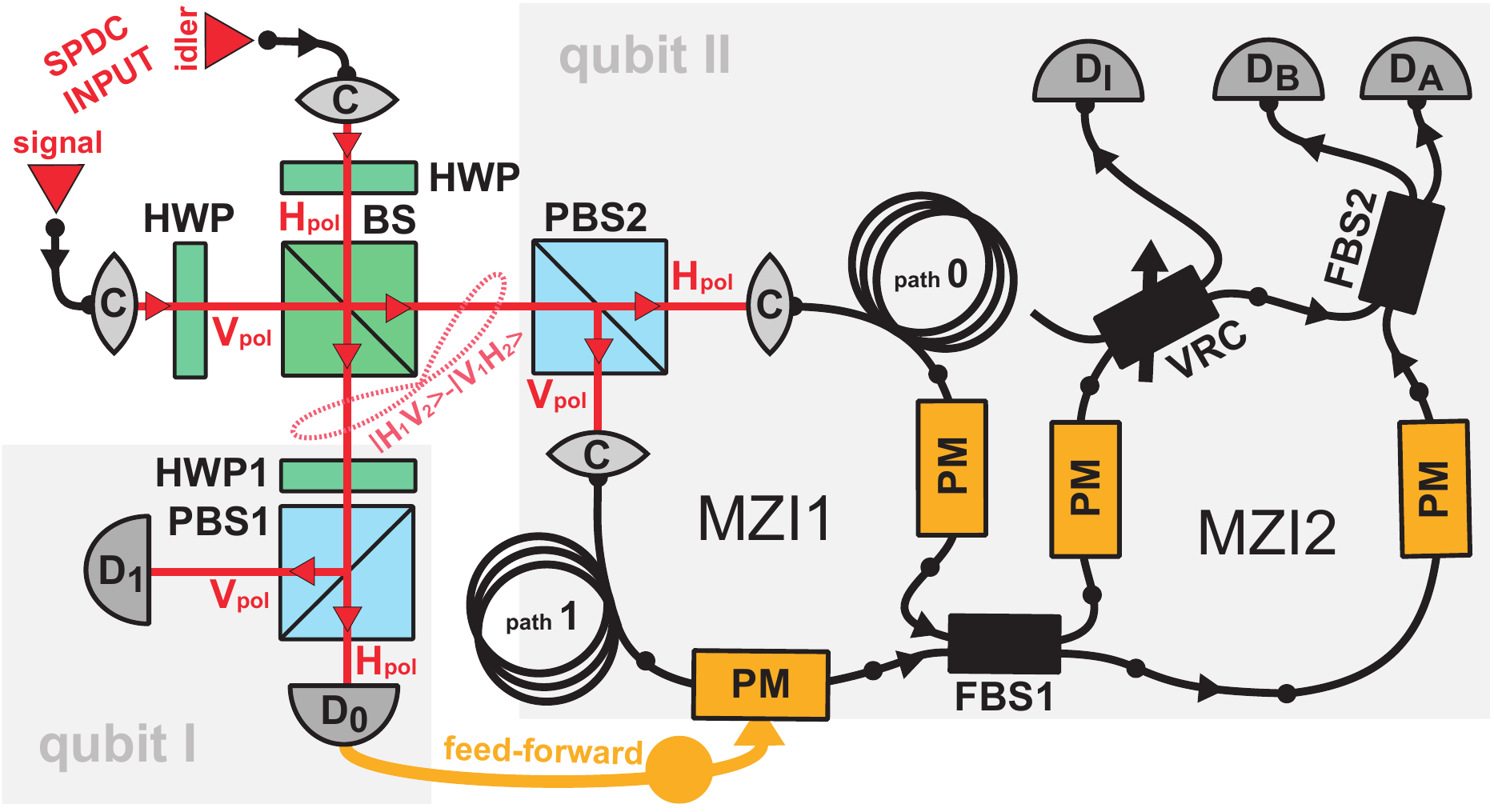}}
  \caption{(Color online) The scheme of the experimental setup, BS - bulk beam splitter 50:50,
FBS - fibre beam splitter 50:50, PBS - polarizing beam splitter,
  HWP - half-wave plate, C - collimating lens, PM - phase modulator, D - single-photon detector. }
\end{figure}

In the main experiment, the measurement that should be identified was performed on the first photon of the entangled pair $|\Psi^{-}\rangle$.
The measurement basis ($\mathcal{M}$ or $\mathcal{N}$) was set by rotating a half-wave plate HWP1 in front of the polarizing beam splitter PBS1.
We associated the basis states $|0\rangle$ and $|1\rangle$ with diagonal $|D\rangle$ and anti-diagonal $|A\rangle$ linear polarizations, respectively.
Namely, $|\phi\rangle=\cos\theta|D\rangle+\sin\theta|A\rangle$ and similarly for other measurement-basis states.
Measurement outcomes $0$ and $1$ were indicated by clicks of detectors $D_0$ and $D_1$, respectively.
Polarization state of the second photon was transformed to path encoding with the help of PBS2 and the photon was coupled into the first of two serially
connected fiber-based Mach-Zehnder interferometers (MZI1). Thus, polarization states
$|V\rangle = (|D\rangle + |A\rangle)/\sqrt{2} \equiv |+\rangle$ and $|H\rangle = (|D\rangle - |A\rangle)/\sqrt{2} \equiv |-\rangle$
were then represented by a photon propagating in the lower and upper interferometer arm, respectively.
We employed polarization maintaining fibers which suppressed unwanted changes of photon's polarization state during its propagation in the fibers.
Both interferometers MZI1 and MZI2 were thermally isolated and actively stabilized to reduce phase drifts caused by temperature fluctuations and air flux.
If detector $D_0$ registered a photon then an electronic feed-forward \cite{feedforw} conditionally changed the state of the second photon
in MZI1 by applying a $\pi$-phase shift in the lower interferometer arm.
This resulted in transformation $|\phi^\perp\rangle \rightarrow |\psi\rangle $ and $|\psi^\perp\rangle \rightarrow |\phi\rangle$
which is equivalent to the conditional application of unitary operation $\sigma_Y$ in Eq. (\ref{sigmaYconnection}) up to an exchange of the role of $|\phi\rangle$ and $|\psi\rangle$.

\begin{figure} [!t!]
\centerline{\includegraphics[width=0.98\linewidth]{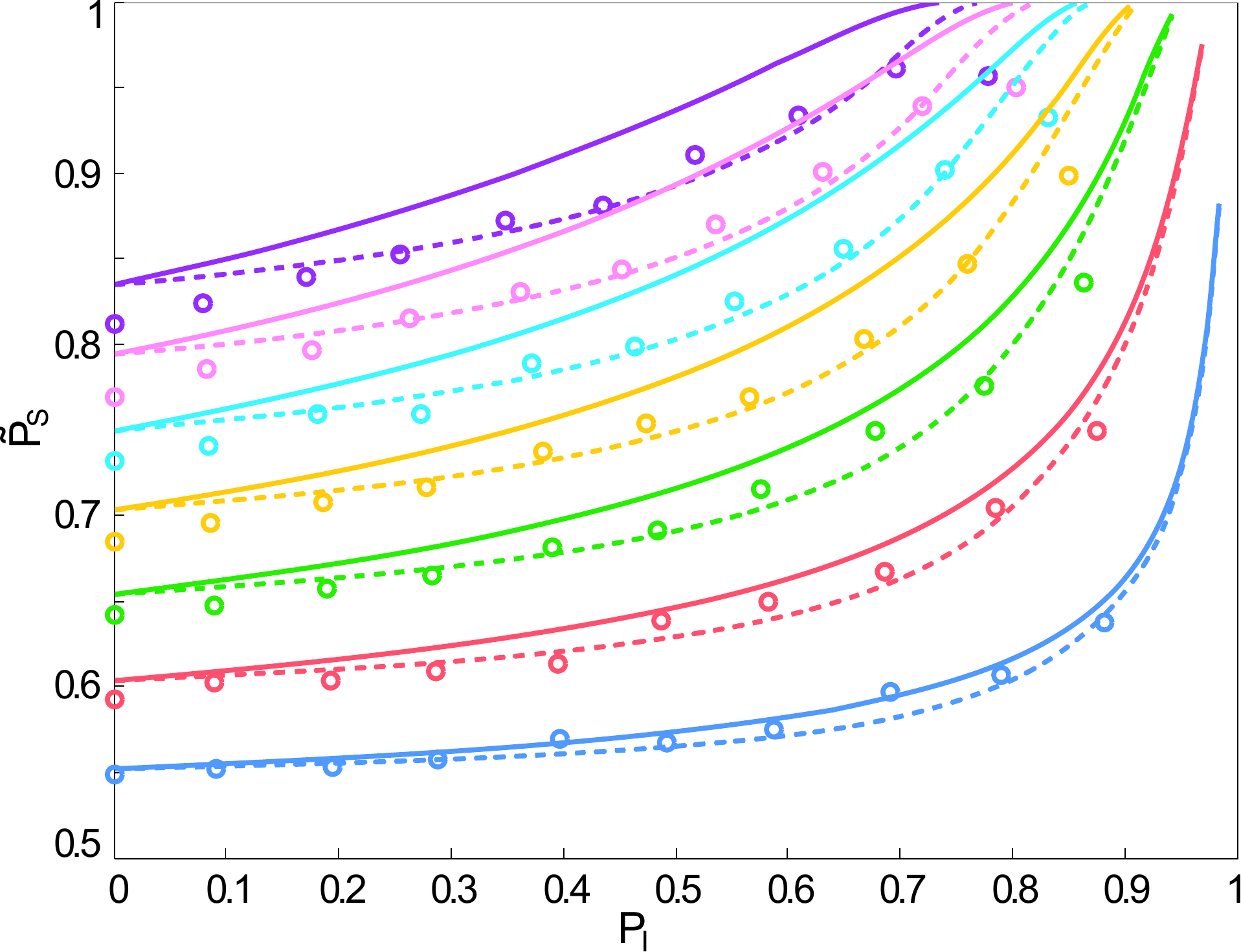}}
  \caption{(Color online) Dependence of relative success probability  $\tilde{P}_S$ on probability of inconclusive results $P_I$ is plotted for $7$ values of $\theta_j=j\pi/30$, $j=1,2,3,4,5,6,7$.
  The value of $j$ increases from bottom to top. Shown are the experimental data (circles) as well as the maximum $\tilde{P}_S$ achievable by the optimal scheme using entangled state (solid lines), and
  using single-qubit probes only (dashed lines).}
\end{figure}

The discrimination problem was thus reduced to a discrimination between two single-qubit states $|\phi\rangle$ and $|\psi\rangle$.
Behind the balanced fiber coupler FBS1 propagation of a photon through the upper (lower) arm corresponded to
the state $|0\rangle$ ($|1\rangle$).
A variable-ratio coupler (VRC) placed in the upper arm of MZI2 was used
as a variable attenuator of the amplitude of the basis state $|0\rangle$, hence it implemented the filter $F$. Projection onto the superposition states $|\pm\rangle$
was achieved using the final balanced fiber coupler FBS2 and detectors $D_A$ and $D_B$.
To determine the probability of inconclusive events, additional detector $D_I$  was used to monitor the output of the tunable fiber coupler VRC. For each basis $X=M,N$ we have measured $6$ two-photon
coincidences $C_{ik}^X$ represented by simultaneous clicks of pairs of detectors $D_i$ and $D_k$, where $i=0,1$, and $k=A,B,I$.
We had measured the relative detection efficiencies $\eta_i$, $\eta_k$ of the detectors, and their influence was compensated
by rescaling the measured coincidence rates as $C_{ik}^X\rightarrow C_{ik}^X/(\eta_i \eta_k)$. The measurement time was the same for both bases
which corresponds to equal a-priori probabilities of $\mathcal{M}$ and $\mathcal{N}$. The probabilities $P_S$ and $P_I$ were then determined as
 $P_{S}=(C_{0A}^M+C_{1B}^M+C_{1A}^N+C_{0B}^N)/C_{\mathrm{tot}}$ and $P_{I}=(C_{0I}^M+C_{1I}^M+C_{0I}^N+C_{1I}^N)/C_{\mathrm{tot}}$, where $C_{\mathrm{tot}}$ denotes the sum of all $12$ measured coincidence rates.

\section{Results}

We have performed measurements for $7$ values of $\theta=j\pi/30$, $j=1,2,3,4,5,6,7$.
For each fixed $\theta$, the transmittance of VRC was varied from $1$ to $0.1$ with the step of $0.1$.
The resulting dependence of $\tilde{P}_S$ on $P_I$ is plotted in Fig.~4 by circles together with the
theoretical curves representing the maximum $\tilde{P}_S$ achievable by the optimal entanglement assisted protocol (solid lines) and by using the single-qubit probes (dashed lines).
The statistical errors of the results are smaller than the size of the symbols. We can see that for certain $\theta$ and $P_I$ the experimental entanglement-based
discrimination indeed outperforms the best strategy without entanglement.  The slight reduction of the experimentally observed $\tilde{P}_S$ with respect to the theoretical prediction could be attributed
to various experimental imperfections such as phase fluctuations inside MZIs, arm disbalance,
slight deviations in phase and polarization settings, slightly unbalanced splitting ratios of beam splitters, and small imperfections in the input singlet state.
As indicated by the theoretical curves, the entanglement-based protocol in theory outperforms the single-qubit scheme for all $P_{I}>0$.
The entanglement thus does not help only in the regime of minimum error discrimination ($P_I=0$) where the optimal success probability $[1+\sin(2\theta)]/2$
can be achieved by a single-qubit probe prepared in state $|+\rangle$. Unambiguous discrimination with single-qubit probe is possible only if the
probe is prepared in a state orthogonal to one of the projectors (\ref{MNdefinition}), say $|\vartheta\rangle=|\psi^\perp\rangle$.
The resulting probability of inconclusive outcomes $P_I=[1+\cos^2(2\theta)]/2$ is larger than the
probability $\cos(2\theta)$ achieved by the entanglement based scheme and the difference increases with $\theta$. We have carried a separate test of unambiguous discrimination for
$11$ different $\theta_j=\arctan(\sqrt{T_j})$ corresponding to transmittances of the VRC, $T_j$, varied from $0$ to $1$ with step $0.1$.
The experimental results, plotted in Fig.~5, are in good agreement with theory and the probability of errors $P_E$ does not exceed $3.2\%$.

\begin{figure} [!b!]
\centerline{\includegraphics[width=0.95\linewidth]{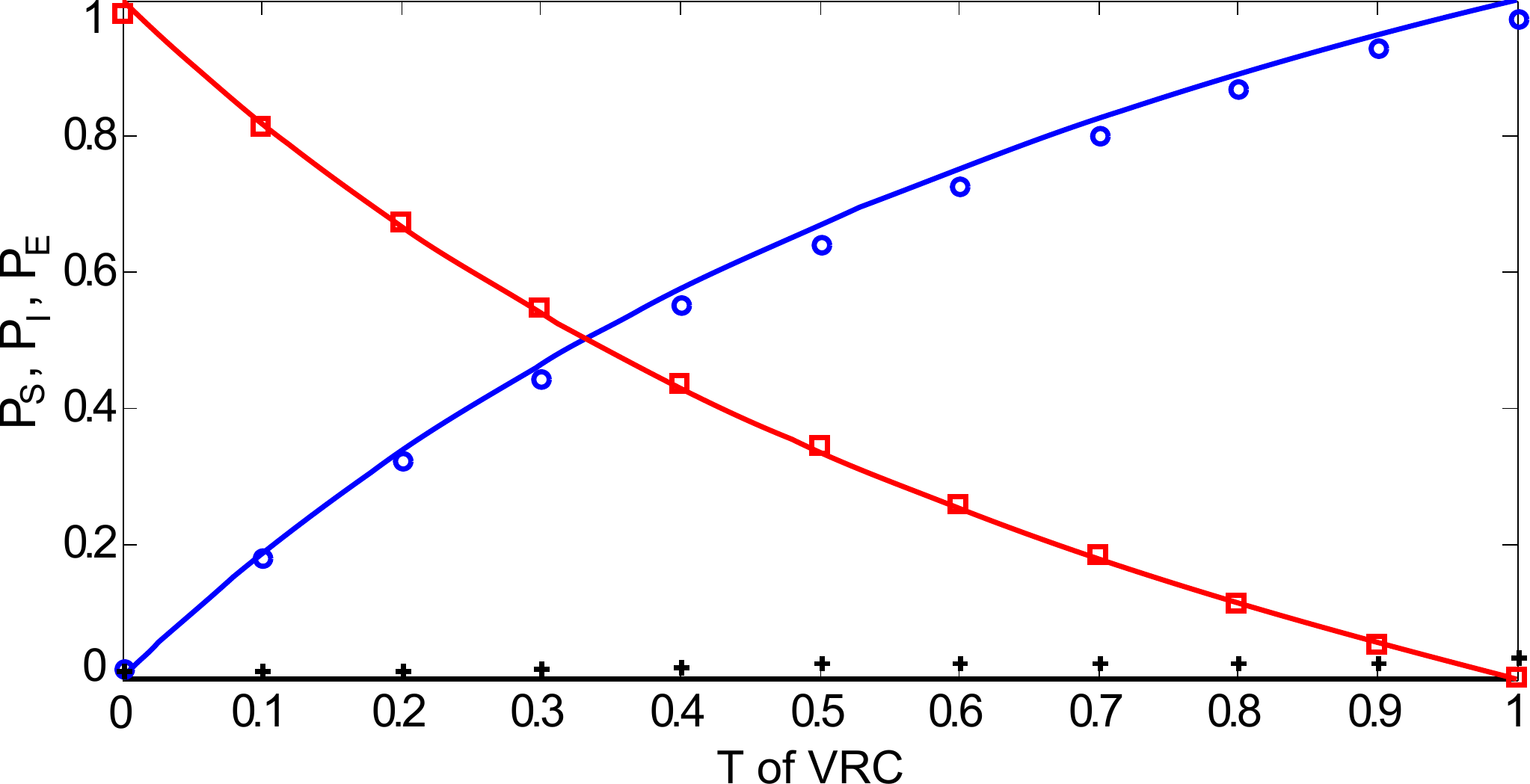}}
  \caption{(Color online) Unambiguous discrimination of quantum measurements. The probabilities $P_S$ (blue circles), $P_I$ (red squares), and $P_E$ (black crosses) are plotted as functions of the VRC splitting ratio $T$.
   The lines represent theoretical predictions. }
\end{figure}

\section{Conclusions}

In summary, we have determined theoretically and implemented experimentally optimal strategies for discrimination between two projective single-qubit
quantum measurements.  The experiment demonstrates that the quantum optical technology is mature enough to harness the benefits of entanglement
in quantum device discrimination, although the entanglement-based scheme is much more demanding than the single-qubit probe scheme,
as the former requires a real-time feed-forward to fully exploit the potential of entangled probes. The techniques and results reported here can be extended to
unequal a-priori probabilities of  $\mathcal{M}$ and $\mathcal{N}$, noisy measurements, and POVMs containing more than 2 elements \cite{mdiscrnew}.
Our findings provide fundamental insight into the structure of optimal probabilistic discrimination schemes for quantum measurements and they pave the way
towards potential applications of such techniques in quantum information science and beyond.

\acknowledgments
This work was supported by the Czech Science Foundation (13-20319S). M.S. acknowledges support by the Operational
Program Education for Competitiveness - European Social Fund
 (project No. CZ.1.07/2.3.00/30.0004) of the Ministry of Education, Youth and Sports of the Czech Republic.
 M.Z. acknowledges the support of projects VEGA 2/0125/13 (QUICOST), COST Action MP 1006 and APVV-0646-10 (COQI).

 \begin{figure}[!b!]
\includegraphics[width=0.98\linewidth]{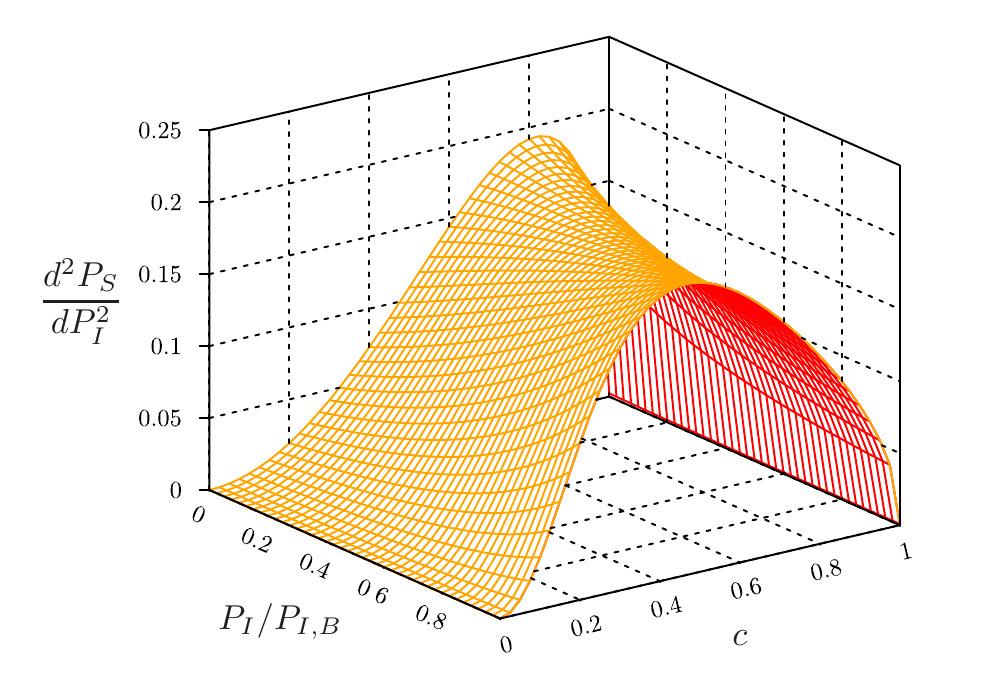}
\caption{The second derivative $d^2 P_S/dP_I^2$ given by Eq. (\ref{PSD2})  is plotted as a function of $P_I$ and $c$.}
\end{figure}

\appendix*

\section{Properties of $P_S$ in the protocol with single-qubit probes}

Here we discuss in detail the properties of the probability of  successful discrimination $P_S$ in a scenario where the two projective
single-qubit measurements are discriminated using one pure single-qubit probe.
In particular, we prove that the success probability $P_S$ given by Eq. (\ref{PS}) is a convex function of $P_I$ on the entire interval $0<P_I<P_{I,B}$, i.e.
\begin{equation}
\frac{d^2 P_S}{d P_I^2}>0.
\label{PSconvexity}
\end{equation}
It is convenient to introduce a new variable $y=cx$. It follows from Eq. (\ref{xcubic}) that $y$ is a root of a cubic equation
\begin{equation}
y^3-2y^2+(1-P_I)y+P_I c^2=0,
\label{ycubic}
\end{equation}
which defines $y$ as an implicit function of $P_I$. If we make the substitution $x=y/c$ in Eq. (\ref{PS}) we get
\begin{equation}
P_S=\frac{1}{2}(1-P_I)+\frac{\sqrt{1-c^2}}{2c}\sqrt{c^2-y^2}\left[1-\frac{P_I}{1-y}\right],
\label{PSabove}
\end{equation}
where $y$ depends on $P_I$ through Eq. (\ref{ycubic}).

After some algebra we arrive at
\begin{equation}
\frac{d^2 P_S}{d P_I^2}=\frac{\sqrt{1-c^2}}{2c}\left(\alpha y'+\beta y'^2+\gamma y''\right),
\label{PSD2}
\end{equation}
where
\begin{equation}
y'=\frac{d y}{d P_I}, \qquad y''=\frac{d^2 y}{d P_I^2},
\end{equation}
and
\begin{eqnarray}
\alpha&=&\frac{2(y-c^2)}{\sqrt{c^2-y^2}(1-y)^2}, \nonumber \\[2mm]
\beta&=& \frac{P_I(3c^2y^2+c^2-2c^4-2y^3)-c^2(1-y)^3}{(c^2-y^2)^{3/2}(1-y)^3}, \nonumber \\[2mm]
\gamma&=&\frac{(y-c^2)P_I}{\sqrt{c^2-y^2}(1-y)^2}-\frac{y}{\sqrt{c^2-y^2}}. \nonumber \\
\end{eqnarray}
The derivatives $y'$ and $y''$ can be determined by repeatedly differentiating Eq. (\ref{ycubic}) with respect to $P_I$, which yields
\begin{equation}
y'=\frac{y-c^2}{3y^2-4y+1-P_I},
\end{equation}
\begin{equation}
y''=2\frac{y'+y'^2(2-3y)}{3y^2-4y+1-P_I}.
\end{equation}
When evaluating the second derivative (\ref{PSD2}), we should use the root of cubic equation (\ref{ycubic}) which maximizes the probability of success (\ref{PSabove}).
 The dependence of $\frac{d^2 P_S}{d P_I^2}$ on $P_I$ and $c$ is plotted in Fig.~6.
We can see that the second derivative is non-negative for all $0\leq c\leq 1$ and $0 \leq P_I \leq P_{I,B}$.

When $P_{I}>P_{I,B}$, then it is optimal to set $q=0$ and $P_S$ is given by Eq. (\ref{PSsingle}), which is a concave function of $P_I$.
In this case the second derivative can be explicitly calculated, and we get
\begin{equation}
\frac{d^2 P_S}{dP_I^2} =-c\sqrt{1-c^2} \left[c^2-(1-2P_I)^2\right]^{-3/2}<0.
\end{equation}

\end{document}